\documentclass{article}

\usepackage{amssymb,amsmath,theorem,euscript,longtable}

\usepackage[cp1251]{inputenc}
\usepackage[english,russian]{babel}

\usepackage{graphicx}
\renewcommand{\refname}{References}

\newcounter{nomer}
\def\nomer{\refstepcounter{nomer} \arabic{nomer}. }

\newcounter{sec}

\newcounter{punct}[sec]

\def\punct{\refstepcounter{punct}{\arabic{sec}.\arabic{punct}.  }}

\def\COUNTERS{\addtocounter{sec}{1}
              \setcounter{punct}{0}
          \setcounter{equation}{0}
          \setcounter{theorem}{0}
                  }

\begin{document}

\renewcommand{\refname}{References}

 \def\ov{\overline}
\def\wt{\widetilde}
 \newcommand{\rk}{\mathop {\mathrm {rk}}\nolimits}
\renewcommand{\Im}{\mathop {\mathrm {Im}}\nolimits}
\renewcommand{\Re}{\mathop {\mathrm {Re}}\nolimits}
\def\sm{\smallskip}

\begin{center}
\bf\Large
On statistical researches of  parliament elections in the Russian Federation, 
04.12.2011 

\bigskip

\sc \large
Yury Neretin
\end{center}

{\small There is a lot of statistical researches of Russian elections 04.12.2011.
The purpose of this activity is to give a mathematical proof of large falsifications and 
to estimate  possible 'real results of elections'. My purpose is to show that

1. Statistical argumentation allows to prove existence of falsifications and 
to give a lower estimate of falsification, near 1 percent (may be, slightly more). 

2. Statistical proofs of stronger
statements are incorrect from both points of view of mathematics and of natural sciences.

3. Apparently,  this problem is not a problem of pure mathematics
(since it includes strong indeterminacy of sociological nature).}

\bigskip

This is  an elementary note about statistical researches of Russian parliament elections 04.12.2011
and about statistical mistakes
({\sf this is not a text  about elections and electoral falsifications!}).
Such kind of re\-sear\-ches with mathematical slogans occupied Russian newspapers and Russian Internet
(as livejournals, etc.), they are  reflected in some Western editions
as  Washington Post\footnote{ 
See \cite{Sim}} and Wall Street Journal.
 There are also mathematical slogans and posters on meetings%
\footnote{See preprint Simkin \cite{Sim}. The  Russian
scientific newspaper 'Troitskij variant'
20.12.2011 \cite{TV} was begun by a banner {\it 'Gauss instead of cobblestone'},
explanation  for a non-Russian reader is given below on Figure \ref{fig:shadr}.}.
%I have to make a remark for a non-Russian reader:
% "Cobblestone is a weapon of proletariat" is the famous sculpture
%of Ivan Shadr, 1927, for a reproduction, see, e.g., 
%{\tt http://sculpture.artyx.ru/books/item/f00/s00/}. 
Texts of such level can not be a subject of scientific reviews or scientific dis\-cus\-sions.
On the other hand such review seems necessary. 

However,
 two similar papers \cite{Shp1}, \cite{Shp2} of Dmitry Shpilkin were  published in two Russian 
 serials of scientific community, the
electronic journal {\it Za-nauku.ru} and the newspaper {\it Troitskij variant}.
This makes reviewing possible%
\footnote{On the history of his argumentation, see \cite{Pet}.}. 

%%%%%%%%%%%%%%%%%%%%%%%%%%%%%%%%%%%%%%%%%%%%%%%%%%%%%%%%%%%%%%%%%%%%%%%%%%%%
%%%%%%%%%%%%%%%%%%%%%%%%%%%%%%%%%%%%%%%%%%%%%%%%%%%%%%%%%%%%%%%%%%%%%%%%%%%%%%
%%%%%%%%%%%%%%%%%%%%%%%%%%%%%%%%%%%%%%%%%%%%%%%%%%%%%%%%%%%%%%%%%%%%%%%%%%%%%%%
%%%%%%%%%%%%%%%%%%%%%%%%%%%%%%%%%%%%%%%%%%%%%%%%%%%%%%%%%%%%%%%%%%%%%%%%%%%%%%%%%%%

\medskip

{\sf
\begin{center}
Contents
\end{center}

1. Samples. 
Peculiar properties of the modern Russian Federation and of the Duma elections, December 2011.

2. Existence theorem  of falsifications.

3. Non-Gaussian distributions.

4. Distribution of voting turnout. 

5. Correlations between voting turnout and results
of voting.

6. Is Moscow metropolis homogeneous? 

7. Final remarks.

References

Addendum 1. Regional statistics.

Addendum 2. Voting stations with given results of United Russia.}

\bigskip

Shpilkin \cite{Shp1}, \cite{Shp2}
produces several statistical distributions (the detailed discussion is below) 
and says that they can not be obtained 
by the usual electoral process. Therefore, these odd distributions detect falsifications. 

\sm

We use contra-arguments of 3 types:

A. Simple arguments of pure mathematics.

B. Experiments. Since we have  methods of detection of electoral falsifications, 
we can apply them to other countries. The results are challenging...  

C. Oddities  of statistical distributions  admit natural alternative
 explanations. Sometimes such explanations are self-obvious,
 sometimes their  verifications  lead to sociological problems.

\sm

Each group A, B, C of reasonings (see below) is a sufficient  counter-argument to  \cite{Shp1}, \cite{Shp2}.

\section{Samples. 
Peculiar properties\\ of the modern Russian Federation \\and of the Duma elections, December 2011}

\COUNTERS

Before a discussion of statistical distributions we need some preliminary  remarks about properties of samples.

\bigskip

{\bf A. Parties} 
 
 \bigskip
 
 {\bf\punct } Parliament ({\it Duma})
  voting is organized according party lists (no voting for individual
  candidates). Each party
 has a federal list of candidates and regional lists. Total number of position of a party in 
 the parliament  
 is determined by the global voting. The final regional distribution
 of par\-lia\-men\-ta\-rians
  depends on regional
election results.  
 
 \sm
 
 {\bf \punct  List of political parties.}
Main parties%
\footnote{Their political orientations
is no matter for this note.} 
(which were present in the previous Duma and will be present in the next Duma) 
 
  UR (the United Russia) --- the party of power; 
 
 CPRF --- the Communist Party of the Russian Federation;
 
 LDPR --- the Liberal-Democratic Party of Russia (party of Zhirinovsky);
 
 JR --- the Just Russia.
 
Additional parties:

the Yabloko - declares itself as a party of 'intelligent'%
\footnote{A specific Russian social stratum (интеллигенция), dictionaries usually give a non-precise
English translation 'intellectual'.};

RC --- the Right Cause;

PR -- the Patriots of Russia.
 
\sm 

{\bf \punct Results of elections.}

UR --- 49.32,

 CPRF --- 19.19,   
 
  LDPR --- 11.67,
  
    JR --- 13.24.
    
     These 4 parties 
will be represented in the parliament. UR gets an absolute majority in the parliament
(238 positions from 450). The remaining parties

Yabloko - 3.43,

  PR - 0.97, 
  
   RC - 0.60
   
  are not represented in the parliament due to
the 5-percent barrier%
\footnote{A party that receive   less than 5\% is not represented  in parliament}.
 
\sm 
 
 {\bf\punct Informal comments.} Existing list of parties does not represent neither social strata,
 nor possible political programs. Proper constituent bodies of op\-po\-si\-tio\-nal parties are small.
 
 Part of electorate considers the op\-po\-si\-tio\-nal parties  as hidden (or visible) hands of the 
 party of power, only CPRF is free from such suspicions.

 \bigskip

{\bf B. Inhomogeneity and exceptional regions}

\bigskip

{\bf\punct }  Russia is an extremely territorially inhomogeneous country (this is a traditional property of 
Russia  during the whole its history).

 Modern economical, social, and demographic situations
strongly vary along the country. The same holds for
the confidence in the local powers,
local influence of oppositional parties, and local oppositional leaders.
 
 Inhomogeneity  preserves on the level of regions (there is difference
 between cities, small towns, villages,
  etc.; even
 in different towns situation can be different).

\sm

{\bf\punct }  Social inequality in the modern Russia is essentially larger than in modern Western countries.

\sm

{\bf\punct} Social situation in Russia was seriously changed during the last two years, 2010-2011. 

\sm

{\bf\punct  Exceptional regions.}
 According the Constitution%
 \footnote{\tt http://www.constitution.ru/en/10003000-04.htm}, the Russian Fe\-de\-ra\-tion contains two types of regions.
The regions of the first kind are called {\it oblast}, {\it kray}%
\footnote{In official English translation of the Constitution
 they are  called 'regions', 'territories'.} 
(55 regions), cities of Federal
importance (2 regions). The regions of
the second kind are national
autonomies, they are named {\it republics} (21 regions), {\it autonomous okrugs}%
\footnote{Official English translation is 'autonomies'.}  (4 regions),
\,{\it auto\-no\-mous oblast} (1 region).

 Below we do not discuss 'autonomous okrugs',
because their populations are small (by their definition). They are extreme
north regions with mining industry. 

 'Republics' are different. Republics with Christian Orthodox title nations
 usually  are similar to 
Russian regions.  Some of the remaining 'republics' 
 are ethnocracies or have strong ethnocratic tendencies, it seems
that their aristoi have strong positions with respect to the central government. Below we use the term 
{\it exceptional regions}%
\footnote{I give a definition of exceptional objects in political terms, however they are distinguished
in the table in the next section, also see  below Figures \ref{fig:vsevse} and
\ref{fig:republics}  \label{foot2}} for such republics.

Chechenia is
a local dictatorship inside Russia.   Other exceptional regions are essentially softer,
 but their powers
are sufficiently strong to organize voting in their own interests.  

\sm

{\bf\punct Results of voting in the exceptional regions.}
Here we present the list of regions (excluding 'natsionalnyj okrugs'), where  the dominant party
'United Russia' received more than 2/3 percents of voices%
\footnote{The complete table of regions is contained in Addendum 1.\label{foot:table}}.
We also present the  voting turnout (column 3). The last  column contains the percent of voters 
for United Russia  with respect to the total amount of electors.
In brackets we write total number of electors (in millions).

%\begin{table}
\begin{tabular}{l c c c}
The Republic of Chechenia   (0.6)    &  99.5& 99.5 & 99.0 % 0.6
\\
The Republic of Mordovia (0.7)       & 91.6& 94.2 & 86,3 % 0.6
\\
The Republic of Dagestan  (1.6)       &  91.4& 91.2& 83.6 %1.3
\\
The Republic of Ingushetia   (0.2)    & 91.0&  86.4& 78.6 % 0.15
\\
The Karachaevo-Cherkesskaya Republic (0.3) &89.8& 93.2& 74.7 % 0.22
\\
The Republic of Tyva   (0.2)          & 86.3 & 85.1 & 73.4 % 0.1
\\
The Republic of Kabardino-Balkaria (0.5) & 81.9& 98.4 &80,6 %0.4
\\
The Republic of Tatarstan (3.0)       & 77.8& 79.5& 61.9 % 1.86
\\
The Republic of Bashkortostan (2.9)   & 70.5& 79.3 & 55.9 %1.62
\\
The Republic of North Osetia (0.5) &  67.9 & 85.8 & 58.3 %0.3
\\
The Tambov oblast (0.9) &            66.7 & 68.3& 45.6
\end{tabular}
%\end{table}

This list contains only two non-exceptional regions,  (agricultural)   Tambov oblast (on the low  boundary
of 2/3
with  relatively small voting turnout) and
the  Republic of Mordovia, whose 'result' seems completely 'mysterious'.
 
 The total percent of voices for United Russia in the whole
 country is 49.3, the voting turnout is 60.2%
 \footnote{We have $49.3\times 60.2= 0.3$. In this sense
the Tambov oblast is more similar to ordinary regions than to the exceptional 'republics'. See Addendum 1.}. 
 The total amount of electors is 109.2 millions.
 
{\it The total number of voices for United Russia was $32.4$ millions.
The exceptional regions (plus the Republic of Mordovia) gave United Russia   $7.3$ millions $(22.5\%)$.}

 \sm
 
{\it The exceptional regions plus the Republic of Mordovia (totally near $10.5$ millions of electors,
near $10$ percents)
  produce numerous exotic properties of electoral  statistical distributions.}
  
  \sm
  
 Consider Russia without the exceptional regions. Then the result of UR is 44.3\%
 (the relative loss of the party is $(49.3-44.3)/49.3$, i.e. 10\% of collected voices).
 Certainly,  without the exceptional regions UR could not receive the absolute ma\-jo\-ri\-ty
 in the parliament.
 
 \sm
 
 It seems that nobody believes in the results of elections in the exceptional regions, however
 such discussion  is not a question of mathematics%
 \footnote{It can happened that real percents of voting for 
 United Russia 
 in some republics were relatively large, but smaller than it was 'drawn'.
 Voting turnout is even more
 impressive.}.
  Local aristoi are interested to send their representatives to the Federal parliament,
   to demonstrate their own power  to local population of regions,
 and to show to the central government both a loyalty and a force.

%32.4

%7.3

%461866+421283+463340+338464+ 665112+703868 + 151257 +428171+ 267475  +565597 + 297704=4764137
%+467867 + 489032+ 541080 + 279210 + 119705+ 607909 +
%7268940

\begin{figure}
\includegraphics[scale=1]{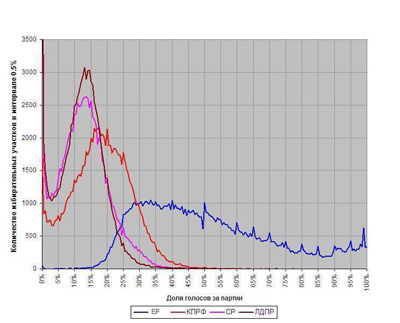}
\caption{From \cite{Shp1}. Station-voting diagram for the whole Russia
The horizontal axis corresponds to percent of voices for a party,
 the vertical axis corresponds to number
of voting station with a given percent. The step is $0.5\%$. Blue curve 
shows the distribution of  UR, red; lilac, brown correspond to
CPRF, JR, LDPR respectively. Integrals $\int f_j(p)dp$ coincide for all graphs
$f_j$.\label{fig:shp1}}.
\end{figure}

 \section{Existence theorem for falsifications}
 
 \COUNTERS
 
{\bf \punct Argument 1. Dents.} D.Shpilkin \cite{Shp1}, \cite{Shp2}
 for 4 main parties evaluates number of voting  stations
with a given percent of voices for a party. He get 4 graphs given on Figure \cite{Shp1}.
We call such figures as {\it station-voting diagrams.}

Certainly, dents on 50, 60, 65, 70, \dots percents is a sufficient argument for existence of falsifications.
It is clear that electoral commissions tried to achieve these 'important' numbers.

Kobak \cite{Kob} claims that dents give lower estimate%
\footnote{This 1\% is partially  due to the exceptional regions.  
Doubtless (by non-mathematical reasoning),
the
exceptional regions produced more than 1 percent...} $\ge 1\%$. 

However, we discuss this more carefully (d'apres Kuznetsov and Kobak)
 
 \sm
 
 {\bf\punct Unexpected obstacle.}
Kuznetsov \cite{Kuz}  presents the same station-voting diagram
  in another resolution, the step is $0.05$ percents.
 
 \begin{figure}
\includegraphics[scale=0.4]{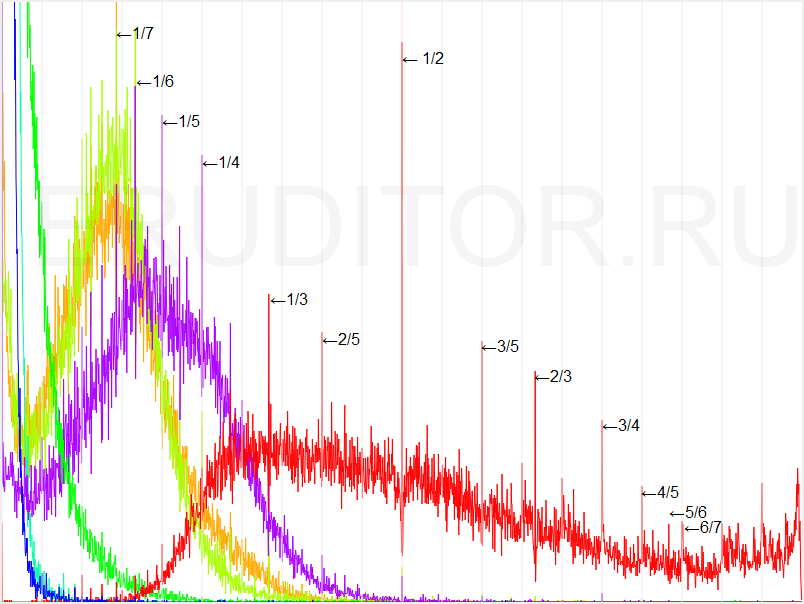} 
\caption{(Kuznetsov \cite{Kuz}) The station-voting diagram (i.e.
Figure \ref{fig:shp1}) with the step $0.05\%$.
Graphs for parties UR (red), CPRF (lilac), JR (yellow), LDPR (green),
Yabloko (dark green), PR (dark blue), RC (blue).
\label{fig:kuz1}}
\end{figure}

\begin{figure}
\includegraphics[scale=0.3]{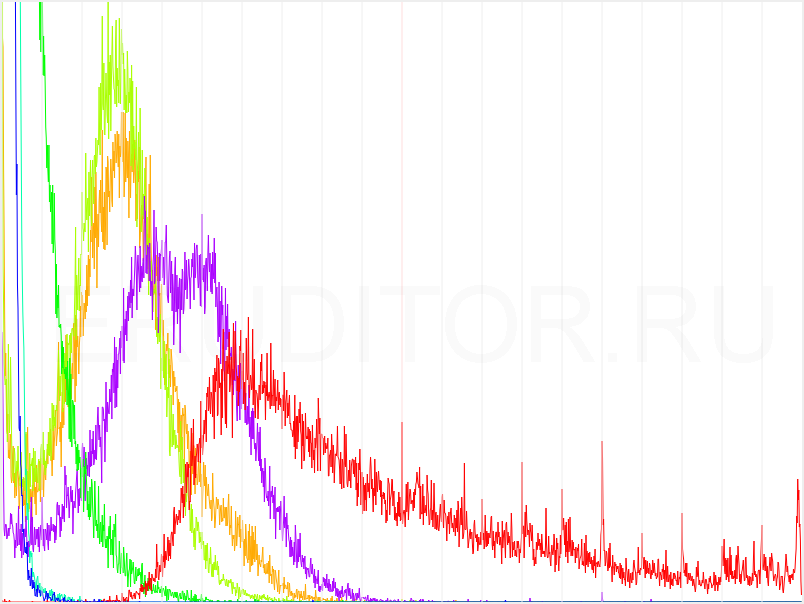} 
\caption{(Kuznetsov \cite{Kuz}) The number of electors on voting stations with
given percent for a given party (this is not a station-voting diagram).
 The resolution is 0.2\%.
\label{fig:kuz2}}
\end{figure}

We observe numerous dents on graphs corresponding to fractions $p/q$ with small numerators
 $q$. 
Apparently, such dents are artifacts
corresponding to small voting stations%
\footnote{Consider a voting  station with $N$ electors. Possible results of a party
is $k/l$ with $l\le N$.  But $1/2$ is repeated among possible fractions approximately
$N/2$ times. On the other hand, 1/2 has a neighborhood free of fractions with denominator
$\le N$. Also, fractions, close to $1/2$ have large denominators.
\newline
 Kuznetsov \cite{Kuz} presents 
results of numerical stimulation of a coin-flip voting on small voting stations. This produces 
sharp-toothed graph instead of expected constant.}.
 Therefore dents at $50$, $60$,  $65\simeq 66.6$, $75$, $80$ on
Figure \ref{fig:shp1} partially correspond to small stations. However, small stations can be easily
eliminated, see Figure \ref{fig:kuz2}%
\footnote{We observe a substantial transformation of the UR-graph.
This shows existence of huge number of small voting stations,
for instance, in villages, in hospitals, in frontier posts, some military objects, 
on ships, in pre-trial detection centers...\label{transform}}. 
 \begin{figure}
\includegraphics[scale=0.5]{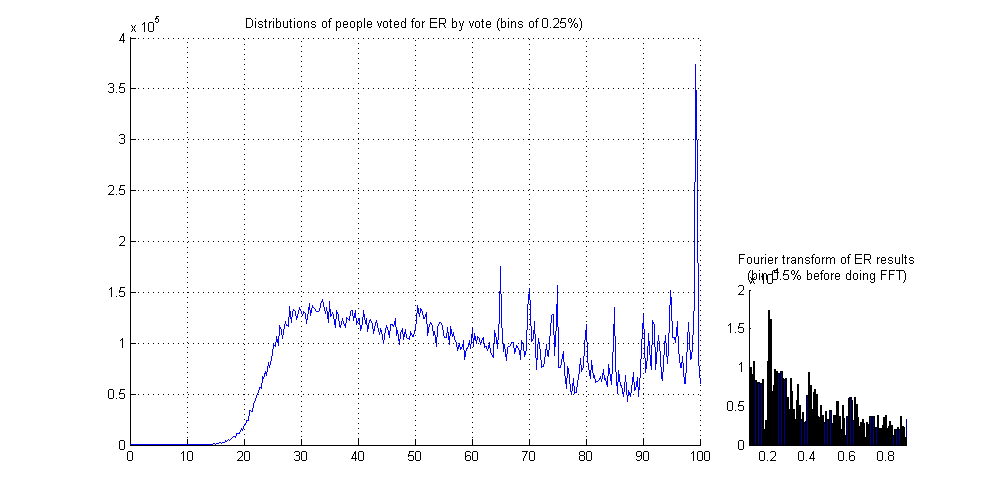} 
\caption{(Kobak) The number of voices obtained UR on electoral stations with
given percent of  voices for UR (this is not a station-voting diagram). The resolution is 0.5\%
\label{fig:kob1}}
\end{figure}
 \begin{figure}
\includegraphics[scale=0.5]{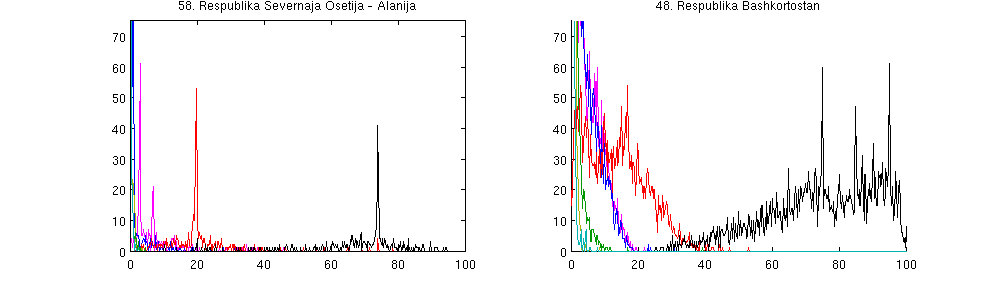} 
\caption{(Kobak) Station-voting diagram  for the Republic of North Osetia and
the Republic of Bashkortostan.
UR is black, CPRF is red.
\label{fig:kob2}}
\end{figure}
Dent at 50\% survive, but the  dent at 60\%
 almost disappears. However, the similar picture with resolution
0.5 percents was given  by Kobak (more precisely, he presents the number of voices for
 UR). Figure \ref{fig:kob1} shows that  60-percent peaks
disappears and 50-percent becomes problematic...

Thus dents at fractions $1/2=0.50$ and $3/5=0.60$ disappear.
However it remains a periodic structure at values 
\begin{multline*}
13/20=0.65,\quad 7/10=0.7,\quad  3/4=0.75,
\quad  4/5=0.80, \\\quad  17/20=0.85,\quad  9/10=0.90,\quad  19/20=0.95
\end{multline*}
Other rational numbers do not produce dents. It seems that it is impossible to invent
an explanation
except 'drawing' of results of elections  or upward rounding%
\footnote{Apparently, it is interesting to draw histograms without small voting stations and histograms
whose steps are not a divisor of 5 percents.}.

To avoid  doubts, we present an impressive  Figure \ref{fig:kob2}
for the Republic of North Osetia and the Republic of Bashkortostan.
Both regions are exceptional in the sense mentioned above.

\sm

{\bf\punct Computer-generated graphics.}
Thus the initial reaction concerning dents was correct. However this discussion shows dangers
hidden in   completely convincing computer graphics.

\section{Non-Gaussian distributions}

\COUNTERS

{\bf\punct The most common argument.} This is: 

\sm

{\sc Observation. }{\it Graphs on  the station-voting diagram
{\rm(}Figure {\rm\ref{fig:shp1})} are not Gaussian!}

\bigskip

{\sc Corollaries:} 

\sm

1) {\it This proves 
the global falsification.}

\sm

2) {\it We can estimate  the real result of UR.} 

\bigskip

This was a topic of posters on meetings (see Figures \ref{fig:meeting1}, \ref{fig:meeting2}),
 this was  multiplied in Internet thousands times,
repeated by numerous journalists,  this came to West newspapers.

\sm

But Gaussian distributions are produced by the Central Limit Theorem, it can not be applied 
to
station-voting diagrams.

\sm

{\bf\punct A  remark.} Let us imagine a Russian region
 voting by a coin flip;  a citizen, which obtains 'head', votes for UR.
In this case, the   
 distribution on the station-voting diagram  is not Gaussian...

Indeed, 
 consider all voting stations  of a fixed size $n$.
The density of the  corresponding distribution is 
$$
\sqrt{\frac {2n}{\pi}}e^{-2n (x-1/2)^2}
$$  
After mixing with respect to $n$
we get a density of the form 
$$
\int \sqrt{\frac {2n}{\pi}}e^{-2n (x-1/2)^2} d\mu(n)
.$$ 
Assume that this density is Gaussian, i.e. $=\sqrt{\frac{2h}{\pi}}e^{-2 h (x-1/2)^2}$.
 We change the variable $y=2(x-1/2)^2$ and write the equation
 $$
\int \sqrt{\frac {2n}{\pi}}e^{-n y} d\mu(n)= \sqrt{\frac{2h}{\pi}}e^{- h y} 
 .$$
Applying $\frac{d^k}{dy^k}$ to the both sides we get
$$
\int \sqrt{\frac {2n}{\pi}}e^{-n y} n^k d\mu(n)= h^k\sqrt{\frac{2h}{\pi}}e^{- h y} 
$$ 
For a fixed $y$ we know all moments of the measure
 $$
d\nu_y(n):= \sqrt{\frac {2n}{\pi}}e^{-n y} d\mu(n)
 .$$
 Therefore, 
 $$
d \nu_y(n)= \sqrt{\frac{2h}{\pi}}e^{- h y} \delta(n-h)
, $$
where $\delta(\cdot)$ is the delta-function.
Therefore, $d \nu_y(n)$ is a delta-measure. Thus, $d\mu(n)$ is a delta-measure.

Sizes of voting stations in Russia strongly vary (from 2-20 electors upto 3000). 
Therefore, {\it for an ideal problem the station-voting diagram is not Gaussian.}

However in this case we get a symmetric one-mode distribution.

\sm

\begin{figure}
\includegraphics[scale=0.3]{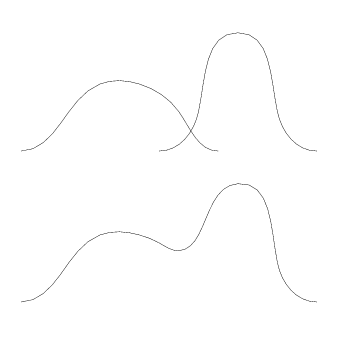} 
\caption{Addition of two Gaussian densities.
\label{fig:wave}}
\end{figure}

{\bf\punct Discussion.} In spite of the previous remark, let us assume that Russia consists
of two ideal regions
with Gaussian distributions on their station-voting diagrams.
 In different regions, these distributions
are different. Therefore we  add two Gaussian densities. We  can get
a  wavy line, see Figure \ref{fig:wave}.

Even if we get a function with one peak, the graph will be asymmetric.

Sum of many Gaussian summands is a graph of non-predictable form. 
For a homogeneous country (as  Sweden) we apparently will get a Gaussian-like picture
(something having one maximum and relatively symmetric). 
For strongly in\-ho\-mo\-ge\-neous Russia there are no reasons to expect a Gauss-like curve.
See also argumentation  in \cite{Sim}.

In note \cite{Ner}, I tried to explain to a reader familiar with modern life of
the Russian Federation
that an  approximate form of graphs of main parties on Figure \ref{fig:shp1} seems to be realistic
(if to keep in mind situation in the exceptional regions).
 Falsifications in ordinary regions deform graphs.
Final results of elections depend on the precise positions of graphs on 
Figure \ref{fig:kuz2}.

\sm

Certainly, if we  omit the exceptional regions from the considerations, then the form of UR-graph changes
(it must be lower at  the segment 65-100\%). But it have to be non-Gaussian.

\sm

\begin{figure}
\includegraphics[scale=0.5]{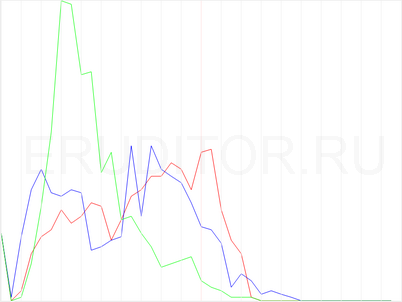} 
\caption{From \cite{Kuz}. 
Elections in Great Britain, 2008. Station-voting diagrams for main parties 
(Conservative, Labor, and Liberal-Democratic).
\label{fig:kuz3}}
\end{figure}

\begin{figure}
\includegraphics[scale=0.3]{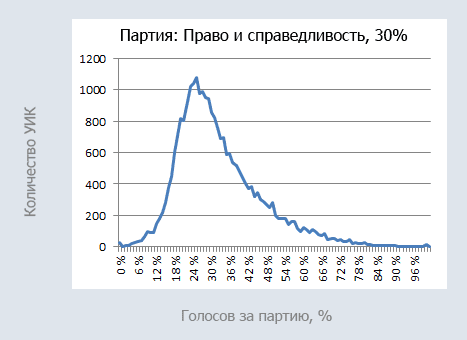}  \includegraphics[scale=0.3]{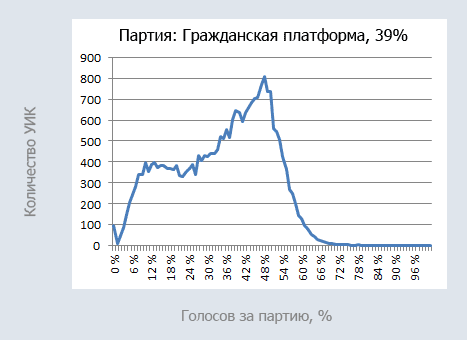} 
\caption{From http://8cinq.livejournal.com/32693.html. 
Parliament elections in Poland, 2011. Station-voting diagrams for  parties 
Prawo i Sprawiedliwosc and
Obywatelska Rzeczypospolitej Polskiej. It seems that Poland is relatively homogeneous.
\label{fig:poland}}
\end{figure}

\begin{figure}
\includegraphics[scale=0.3]{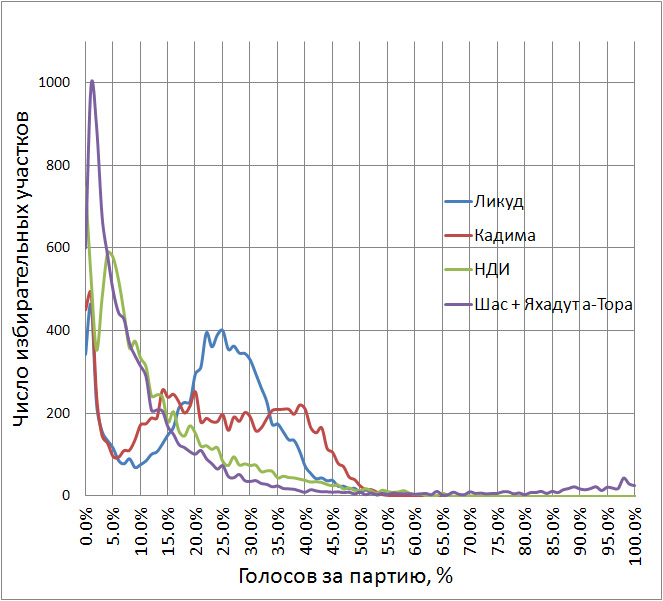} 
\caption{From {\tt http://levrrr.livejournal.com/31427.html}. 
Elections in Israel, 2009. Station-voting diagrams for  parties
Likud, Kadima, Yisrael Beiteinu, Shas+United Torah Judaism.
\label{fig:israel1}}
\end{figure}

{\bf\punct A station-voting distributions Gaussian in West countries?} 
 This  was verified by several authors.
 Kuznetsov \cite{Kuz} presents station-voting diagrams for elections
in Great Britain, 2010 (Figure \ref{fig:kuz3}),
 Kalinin \cite{Kal}  for Canada,
blogger '8cinq' for  Poland (Figure \ref{fig:poland}), blogger 'levrrr' for Israel,
(Figure \ref{fig:israel1}). Graphs, which have no reasons to be Gaussian are not Gaussian. No mystery.

\sm

{\bf 3.5. Computer-generated graphics again.} It is interesting to compare Figures \ref{fig:shp1}
and \ref{fig:kuz2}. Both distributions for UR are non-Gaussian. But  the second figure
is less impressive%
\footnote{... and corresponds to real distribution of voices.},
and only the first  picture
can serve turn a meeting poster (see Figure \ref{fig:meeting1}).

 A transformation of CPRF also it is interesting. On Figure \ref{fig:shp1} the graph has two modes,
 on Figure \ref{fig:kuz2} it has three modes.

%%%%%%%%%%%%%%%%%%%%%%%%%%%%%%%%%%%%%%%%%%%%%%%%%%%%%%%%%%%%%%%%%%%%%

\section{Distributions of voting turnout}

\COUNTERS

\begin{figure}
\includegraphics[scale=0.8]{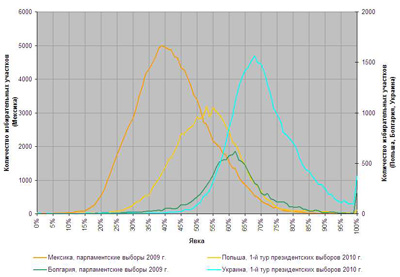}
\caption{From Shpilkin \cite{Shp1}.
Distribution of turnout for 
\newline 
 Mexico, president elections 2010 (brown),
\newline 
 Bulgaria, 2010 parliament elections (green),
 \newline
  Poland,  first tour of president elections 2010
 president elections 2010 (yellow)
\newline
  Ukraine,  first tour of president elections 2010
  (blue)
\label{fig:shp3}}.
\end{figure}

\begin{figure}
\includegraphics[scale=0.4]{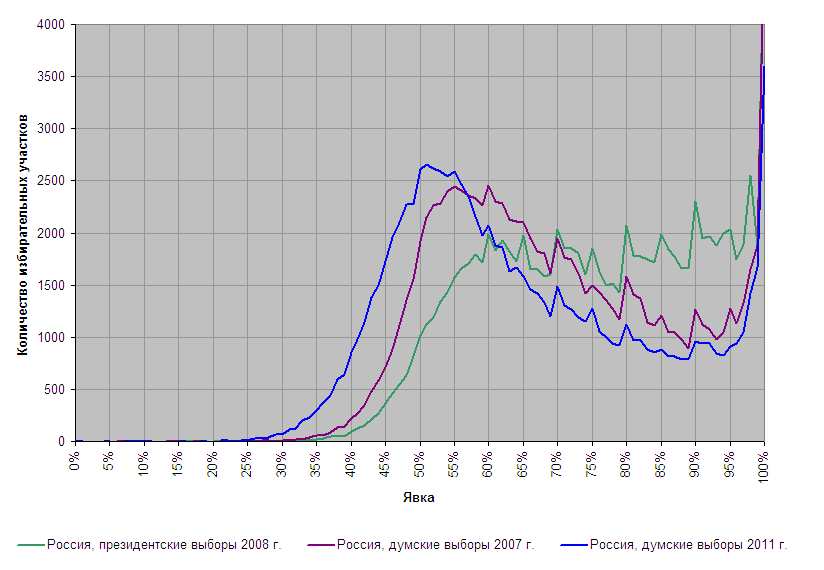}
\caption{From Shpilkin \cite{Shp1}.
Distribution of turnout for elections in Russia:
\newline
president elections, 2008 (green);
\newline
parliament elections, 2007 (lilac);
\newline
parliament elections, 2011 (blue).
\label{fig:shp4}}.
\end{figure}

{\bf\punct Distribution of turnout.}
Next, D.Shpilkin draws diagrams with distribution of turnout, i.e. he evaluates
number of stations with a given turnout. He claims that the distributions in democratic
countries are Gaussian (Figure \ref{fig:shp3}) and in Russia they are  not Gaussian
(Figure \ref{fig:shp4}). 

\begin{figure}
\includegraphics[scale=0.4]{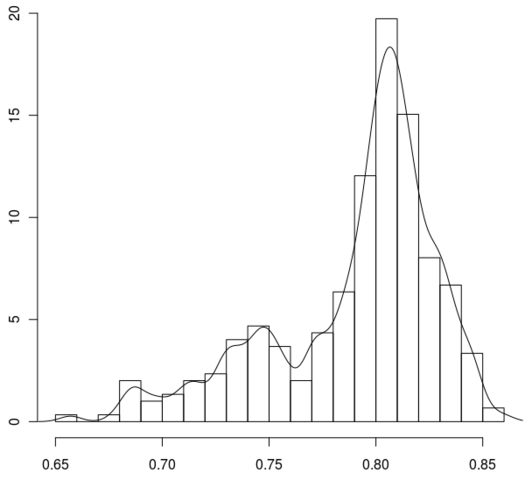}
\caption{From {\tt http://jemmybutton.livejournal.com/1638.html}.
Distribution of turnout for elections in Germany, 2002.
\label{fig:jemmy}}.
\end{figure}

\begin{figure}
\includegraphics[scale=0.4]{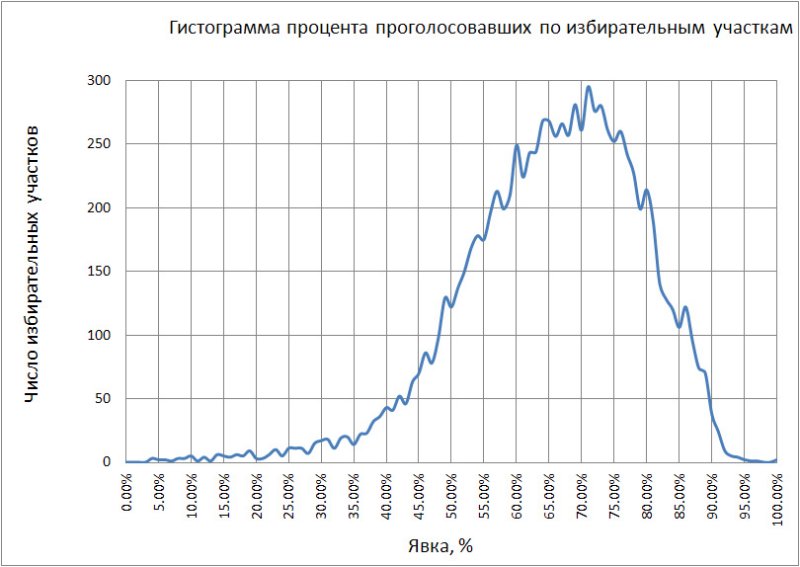}
\caption{From {\tt http://levrrr.livejournal.com/31427.html}.
Distribution of turnout for elections in Israel, 2009.
\label{fig:israel6}}.
\end{figure}

Here we repeat the same arguments as above

1. This distribution has no reasons to be Gaussian in an inhomogeneous country.

2. Assume that electors solve the question 'to vote or not to vote?' by a coin flip.
As it was mentioned above, the corresponding distribution is not Gaussian.  

\sm

{\bf\punct Some West countries.}
We  present graphs  of such distributions for Germany (Figure \ref{fig:jemmy}) and Israel
(Figure \ref{fig:israel6}).

\sm

{\bf\punct Natural explanations.} There are relatively obvious reasons for non-democratic
 parts of Russian graphs  
on Figure \ref{fig:shp4}. For instance:

1) As it was mentioned in Footnote \ref{transform}, there is a lot of small voting stations
in Russia. Such stations easily produce large turnout (see the list in the footnote).
In some cases lists of electors are generated during the voting.

2) Voting in army.

3) Voting in the exceptional regions.

It seems that after removing these components we must get something similar to
(non-Gaussian) Israel graph (modulo a reflection).

\begin{figure}
\includegraphics[scale=0.23]{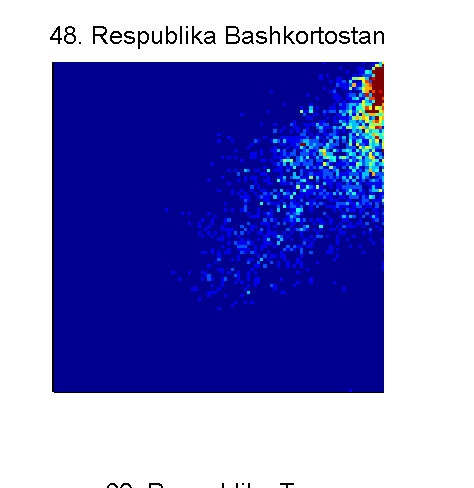}
\includegraphics[scale=0.23]{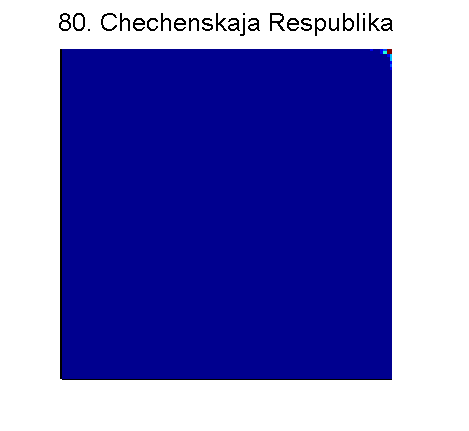}
\includegraphics[scale=0.23]{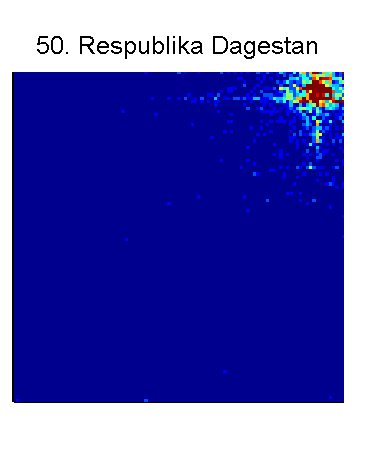}
\\
\includegraphics[scale=0.23]{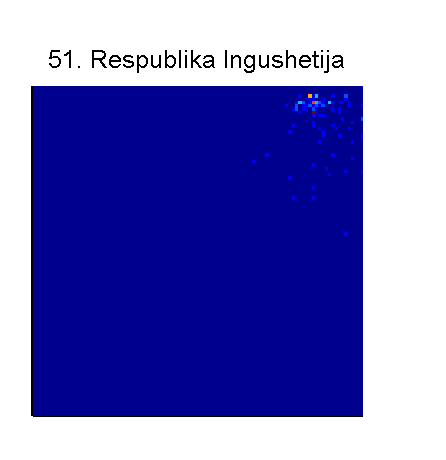}
\includegraphics[scale=0.23]{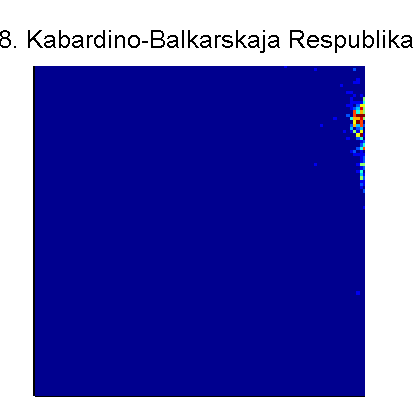}
\includegraphics[scale=0.23]{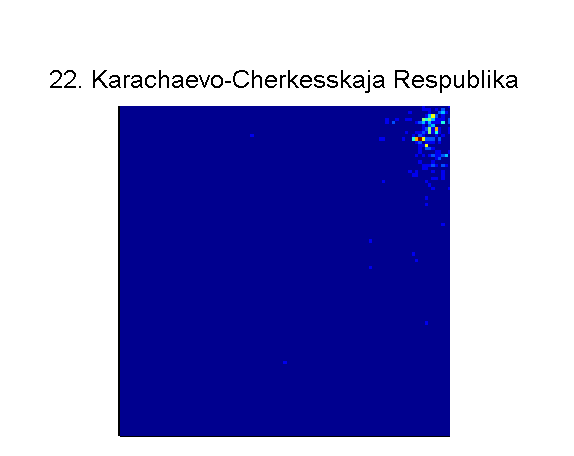}
\\
\includegraphics[scale=0.23]{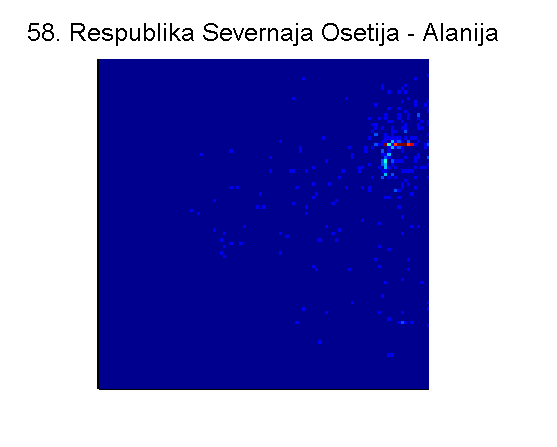}
\includegraphics[scale=0.23]{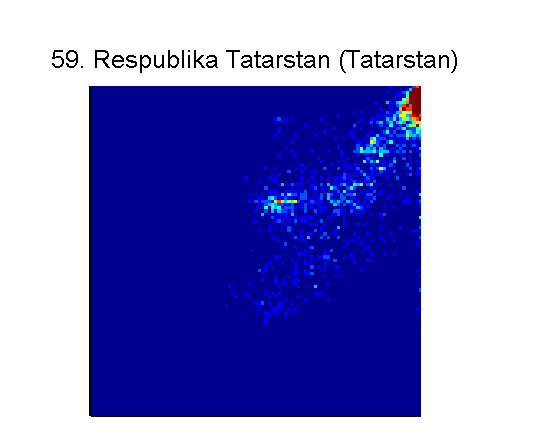}
\includegraphics[scale=0.23]{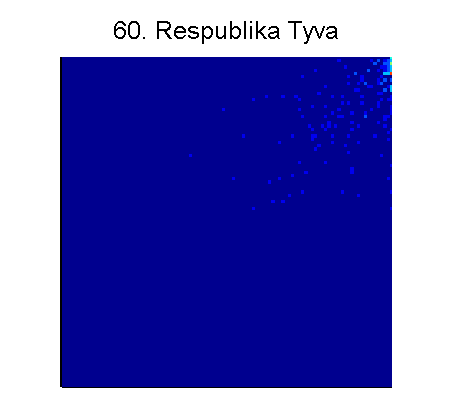}
\caption{UR-clouds in the exceptional regions
\label{fig:republics}}
\end{figure}

Notice also that dents on Figure \ref{fig:shp4} are partially generated by falsifications%
\footnote
{See Figure \ref{fig:republics} below
for the Republic of Daghestan and the Republic of North Osetia, both regions are exceptional.}.
 But sometimes this can be a result of efforts of commission
to collect electors.

\section{Correlations between voting turnout\\ and results of voting}

\COUNTERS

 \begin{figure}
\includegraphics[scale=0.5]{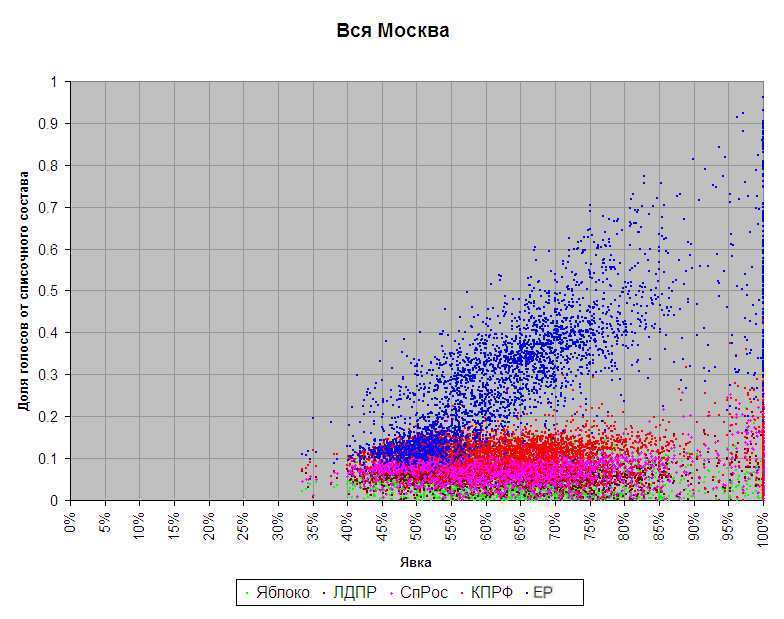} 
\caption{From \cite{Shp1}. The {\it compressed  cloud diagram}.
Elections in Moscow, 2011. For each voting station we 
draw five points, blue for UR, lilac for JR, red for CPRF, brown for LDPR,  green for Yabloko.
The horizontal axis is a voting turnout,  the vertical axis is a percent for a party from the total number
of electors.
\label{fig:shp2}}
\end{figure}

{\bf\punct  Cloud diagrams.}
In computer-generated graphics of Russian elections diagrams of the following two types are popular.

\sm

a) {\it Cloud diagrams.}  For each voting station we draw a point  
$$
(x,y)=
\text{(percent of voting turnout, percent for UR)}
.
$$
Thus we get a cloud of points in the square $0\le x\le 100$, $0\le y \le 100$.

\sm

b) {\it Compressed cloud diagrams.} 
For each voting  station we draw a  point with
coordinates 
\begin{multline*}
(u,v)=\\=
\text{(percent of voting turnout, percent for UR from total number of electors)}
.\end{multline*}

The compressed cloud diagram is obtained from the cloud diagram by the transformation
\begin{equation}
u=x \qquad v=xy
.
\label{eq:nonlinear}
\end{equation}
It is contained in the triangle $0\le y\le x\le 100$.

\sm

{\bf\punct  Correlation.}
D. Shpilkin
\cite{Shp1}, \cite{Shp2} presents the compressed cloud diagram for Moscow, see Figure
\ref{fig:shp2}.

We observe that clouds of 3 main oppositional parties are horizontal,
the cloud for UR is inclined. D.Shpilkin says that the inclination of
UR-cloud is generated by falsifications. He  concludes that only the piece of UR-cloud in the corner
$x\le 55\%$, $y\le 0.2$ is realistic and the rest is falsified., 

In particular, this means a global injection of ballots on large majority of voting stations of Moscow.

\sm

{\bf\punct Doubts-1. West analogs.} The argumentation is based on the axiom:

{\it 'Result of voting on a given station and  voting turnout are independent variables'}.

\sm

\begin{figure}
\includegraphics[scale=0.4]{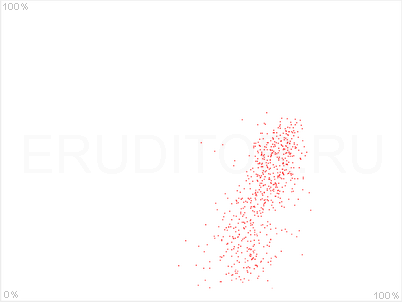} \includegraphics[scale=0.4]{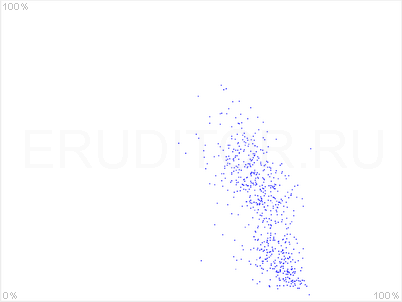} 
\caption{From \cite{Kuz}. Cloud diagram for voting in Great Britain, 2010. The
blue cloud corresponds to the
Conservative Party, the red cloud to the Labor Party.
\label{fig:conlab}}
\end{figure}

\begin{figure}
\includegraphics[scale=0.4]{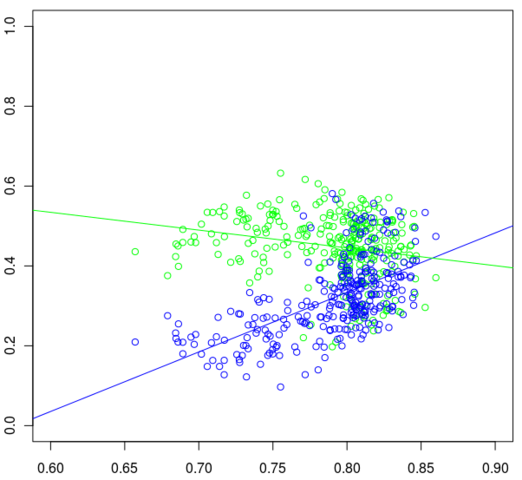}
\caption{Cloude diagram from http://jemmybutton.livejournal.com, 09.12.2011.
 Elections in Germany, 2002.
Blue circles correspond to Christian Democratic Party  + Christian Social Party. Green circles -
Social Democratic party $+$ coalition.
\label{fig:germany}} 
\end{figure}

\begin{figure}
\includegraphics[scale=0.2]{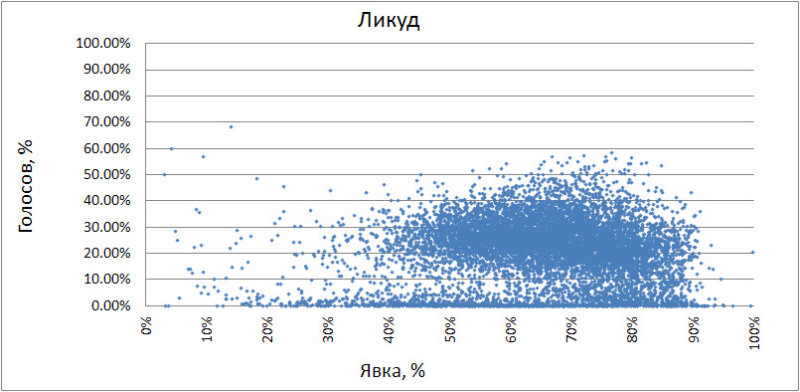}
\includegraphics[scale=0.2]{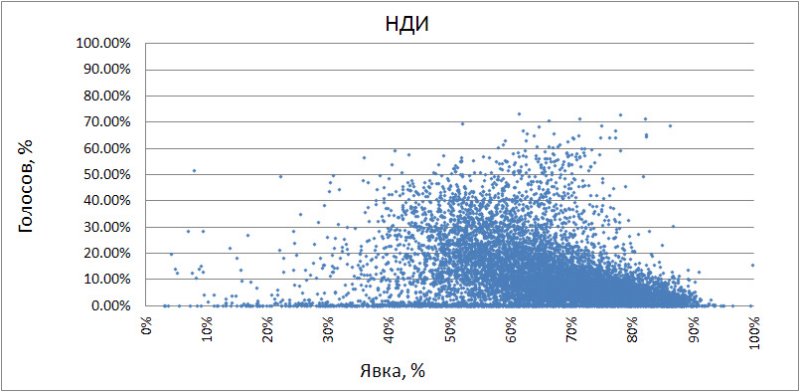}

\includegraphics[scale=0.2]{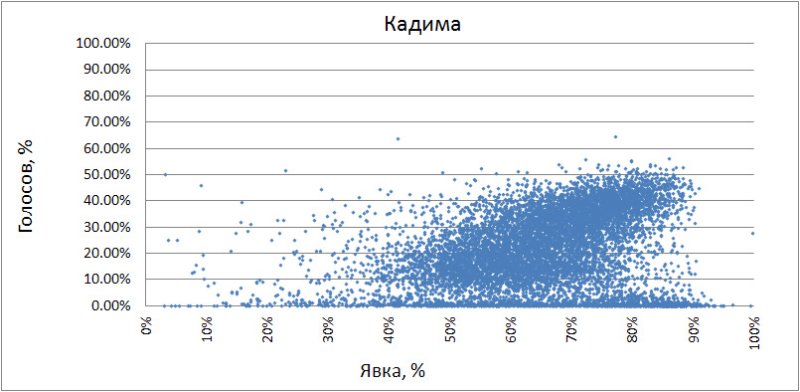}
\includegraphics[scale=0.2]{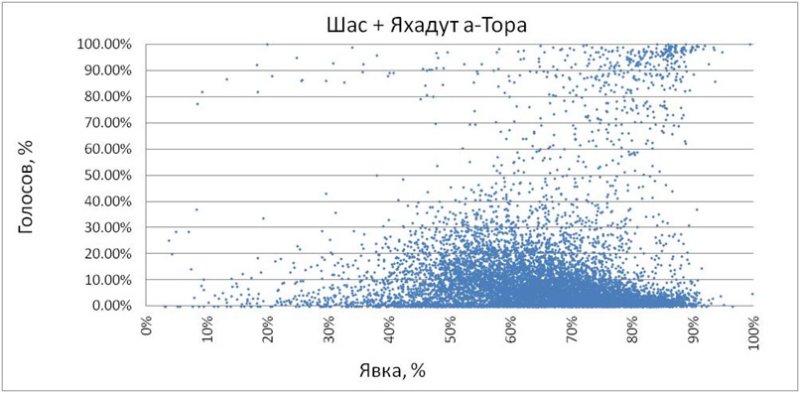}
\caption{From http://levrrr.livejournal.com/31427.html.
 Elections in Israel, 2009.
Cloud diagrams for main parties.
\label{fig:kadima}} 
\end{figure}

 Kuznetsov \cite{Kuz} presents cloud diagrams for elections in Great Britain,
Figure \ref{fig:conlab}. Blogger 'jemmybutton' also checked elections in Germany, 
Figure \ref{fig:germany}, blogger 'levrrr' for Israel
(Figure \ref{fig:kadima}). We observe that the axiom contradicts to ex\-pe\-ri\-ments...

\sm

{\bf\punct Doubts-2. Social reasons.} Accepting the claim of \cite{Shp1}, \cite{Shp2}
we get a global injection of ballots 
on the large majority of voting stations of Moscow. The statement seems suspiciously
strong. 

First, each case of such falsification can be easily unmasked by tools of the usual law
and the usual forensics.
The number of voices in a final report must correspond to the number of signatures
in account books (containing also passport data of each person and the address of his registration).
Therefore it is necessary to write huge number of false signatures in each voting station.
I admit that  statement of \cite{Shp1}, \cite{Shp2}
  can be correct, but in such case this can be
 easily verified  without mathematics
(for more detailed discussion of types of electoral falsifications, see Petrov, \cite{Pet},
see also Buzin, Kynev \cite{Buz}).

Second. Participants of electoral commissions are not members of a secrete guard, they are ordinary people
fulfilling social duties. Lot of voting stations are traditionally located in schools,
many thousands
of Moscow school teachers  work on elections... On the other hand, voting of patients in hospitals
was organized in  hospitals by doctors.%
\footnote{Voting in hospitals produces many voting stations with small number of electors and with large voting turnout.}%
$^,$%
 \footnote{It seems that both strata, school teachers and health professionals, have no reasons
 to like the power of the country.
The best hated ministers in Russia in 2011 were the Minister of  education
 and the Minister of health.}.

\sm

{\bf\punct Doubts-3. On alternative conjectures.}  The simplest conjecture is the following.
Assume that a voting area is homogeneous.
People, who are against  the power, go to elections and vote for one of oppositional parties.
The remains (for instance people, who consider the existing power as lesser evil) do not want too
much to go to elections. If they go, then they vote for UR... In an inhomogeneous area a density
of oppositional population can be a function of point. 

\sm

Another version of the same  conjecture: we have natural voting + falsification. 
In this case we can not separate two parts of the blue cloud.

It can have other reasons.
 However Shpilkin \cite{Shp1}, \cite{Shp2} claims an axiom about independence
 (see above)
  and proposes to accept it
 or to search contra-arguments%
 \footnote{As two axiom about Gauss discussed above (see also an axiom of homogeneity of Moscow
 discussed below).}.

\sm

{\bf\punct Computer-generated graphics.}
I present a quote from   \cite{Shp1}. 

\begin{quote}
'The remaining points of voting for UR are spread
as a diagonal cloud corresponding artificial  raising of voices for UR, and probably,
 lowering of voices for other parties.'
\end{quote} 
 
 Let us notice that scales on the horizontal and vertical axes on Figure
 \ref{fig:shp2} are different... Figure \ref{fig:shp2} implies (if we accept the proposed 
 paradigm) a stronger claim. We continue the discussion in the next section.
 
 \section{Is  Moscow metropolis homogeneous?}
 
\COUNTERS

The topic of the following discussions is not contained in 
Shpilkin \cite{Shp1}, \cite{Shp2}. However
it is a continuation of the previous section and it is one of main arguments of 'electoral science'.

\sm

\begin{figure}
\includegraphics[scale=2]{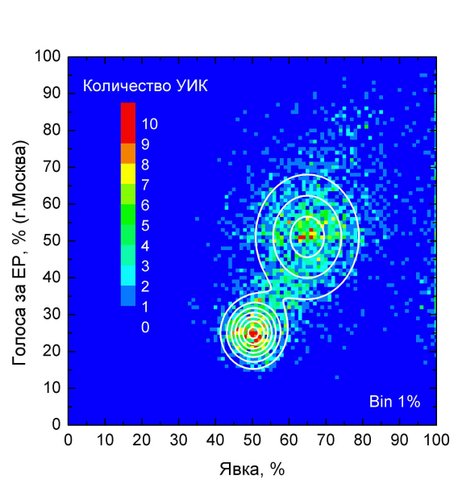}
\caption{From {\tt http://oude-rus.livejournal.com/540865.html}. The UR-cloud from Figure \ref{fig:shp2}. 
Moscow.
Scales on axes are equal.
\label{fig:edro}}
\end{figure}

{\bf\punct Moscow UR-cloud again.} Next, we look to Figure \ref{fig:edro}.
\begin{figure}
\includegraphics[scale=0.25]{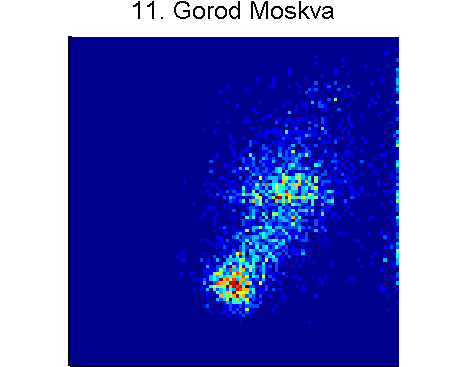}
\caption{(Kobak) The same diagram (I do not understand meaning
of a connected contour on the previous figure).}
\end{figure}
Modas of the UR cloud are at the ponts $(50;25)$, $(65;50)$.
Passing to the compressed diagram we get points
$(50;12.5)$, $(65;32.5)$. The slope is
$$
\frac{32.5-12.5}
{65-50}=\frac {20}{15}=1.33>1
$$
and we observe that the inclination of the ER-cloud is more than 45 degrees.
Hence the result of elections can not be explained by the injection of ballots...

\begin{figure}
\includegraphics[scale=2]{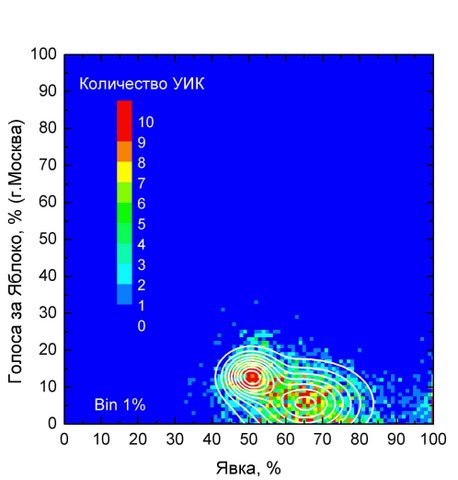}
\caption{From {\tt http://oude-rus.livejournal.com/540865.html}. 
The Yabloko-cloud. Moscow.}
\label{fig:yabloko}
\end{figure}

There is  a 
  conjecture about robbing of the Yabloko (it is  popular among adherents of Yabloko).
   We look to Figure
 \ref{fig:yabloko}. 
First, we observe that there is a correlations between clouds  of UR and of the Yabloko (modas are
on the same vertical lines). 
Modes of the Yabloko cloud are at points $(50,12)$, $(65,5)$. Passing to compressed diagrams
we get points $(50;6)$, $(65;3.25)$.
The inclination is 
$$
\frac{6-3.25}{65-50}=\frac{2.75}{15}=0.18<0.33
$$
 It seems that this is not sufficient to  'help' 
to the United Russia.

Keeping the paradigm we have to assume that the results of elections on the  majority
of Moscow voting stations were more-or-less 'drawn' by electoral commissions%
\footnote{Such commissions in Moscow consist of tens thousands of more-or-less ordinary
people and include numerous representatives of opposition parties. There are to much criminals
(near 15-20 thousands) and they
are too silent  ... This is not
reductio ad absurdum, but we get very strong corollaries from our axioms. 
Such social phenomena must be observable without computer-generated graphics.
}$^,$

\sm

{\bf\punct Is Moscow homogeneous?} In my opinion, this is the first question, which must be asked
by a mathematician looking to Figure \ref{fig:edro}. An important thesis of adherents of electoral
computer graphics is {\it 'Moscow is homogeneous'}. Let us discuss this using simple arguments.

 The next Figure \ref{fig:nice}
 shows a territorial distribution of 
voting for UR in Moscow.  

 \begin{figure}
\includegraphics[scale=0.2]{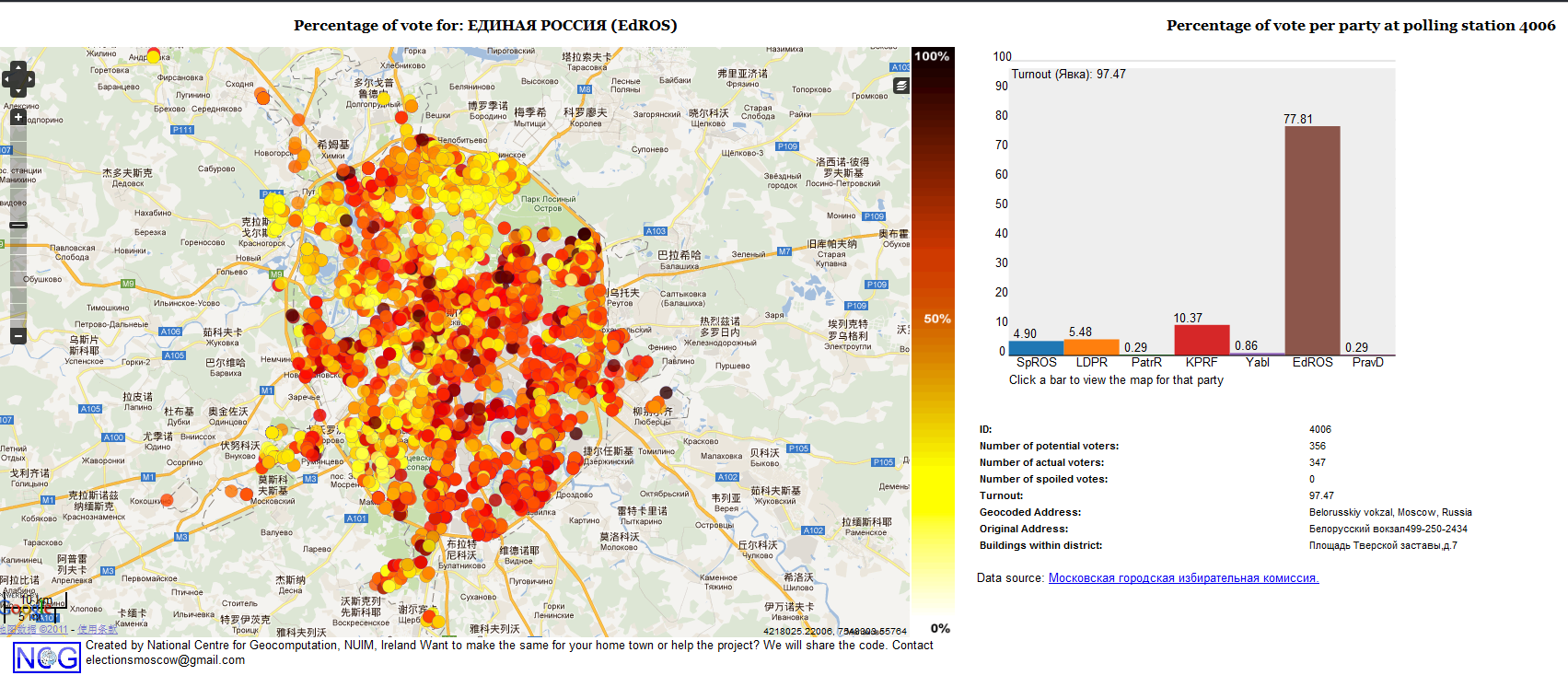}
\caption{From
{\tt http://pics.livejournal.com/kireev/pic/00042rt0}
Territorial distribution of voting for UR in Moscow.
The caption to the map is visible after zoom. Yellow color marks low results of UR, 
red color high result.
\label{fig:nice}}
\end{figure}

Being in the  paradigm of \cite{Shp1}, \cite{Shp2}, we have to assume that 'red' areas
 of Moscow are areas of total falsification,
 'yellow' areas are bastions of honey. 

But it can happened that 'yellow' areas are simply  places of concentration of stratas 
of Moscow population that {\it now} are  'angry'
 (including 'intelligent%
 \footnote{This explains correlation between modes of the United Russia and the Yabloko}' 
 and some other strata of 'middle class'%
 \footnote{Notice that there were two diverse electoral slogans of 'angry' stratas,
 '{\it Vote against the United Russia, the Party of swindlers and thiefs}'
 and '{\it Boycott elections!}'. For this reason,  'angry' population was not sufficiently well-represented
 on elections.}).

\sm

{\bf\punct Inhomogeneity of Moscow population. Soviet time.} The author of these notes
during 20 years (1966-1987) lived in a working class area. Such areas arose in Moscow
in the period of industrialization  and produced usual problems related to such areas.
I only mention high level of aggression of teenagers and tense atmosphere in schools. 
This 'defended' such areas from an infiltration of 'intelligent'.

Later (70-80s) so-called 'bedroom districts' became usual in Moscow. This means that the majority 
of population of these areas were employed far from their houses. Apartments in Soviet time were
distributed by the state. But  the 'state'  in this case was not an abstraction, it was 
represented by numerous economical and social 
 structures of different level, starting with some strong firms 
(as big factories, technological-research centers, etc...)
and ending with ministries.  Such centralized settlement had to produce numerous inhomogeneities
of the population.

On the other hand it was possible to buy a flat with payment by installment ('cooperative').
 Such
projects were initiated by state structures, this produced exceptional buildings
or groups of buildings.

\sm

{\bf\punct Post-Soviet time.} After 1991 Moscow met the usual problems of modern huge cities
(may be, in more rigid forms), it seems that the  growth of non-homogeneity is one of such  
phenomena...

\sm

In any case, Moscow has sufficient reasons to be inhomogeneous, and
 the statement 'Moscow is homogeneous' is non-justified.
In particular, this statement claims that Moscow is an extraordinary metropolis...

Certainly, {\sf the problem of social and ethnic inhomogeneity of Moscow 
 is not a problem of mathematics.}

\section{Final remarks}

\COUNTERS

{\sf I recall that his note is not a text about parliament elections in Russia.}

\sm

{\bf\punct}
I  do not review numerous non-Gaussian electoral
 distributions, which have no a priori reasons to be Gaussian.
 
 \sm
 
{\bf\punct}
There exists also argumentation related to difference between results of voting on stations
equipped with  systems of automatic  counting of ballots and remaining stations. See comments
 in the  papers of Petrov \cite{Pet} and Kobak \cite{Kob}.  

\bigskip

{\bf\punct} Modern methods of investigation of Russian elections  allow to give a mathematical proof
of falsification
of elections in arbitrary sufficiently in\-ho\-mo\-ge\-neous country%
\footnote{Examples mentioned above are Great Britain, Germany, Canada, and (unawares) Poland, Israel.}.
 In particular,
they allow to prove falsification of any elections, which can happen  in  Russia
(and apparently to assign any desired percent of falsification).

\sm
  
{\bf\punct} Tens of men equipped with modern computers can produce numerous
 correlations that can be declared
as impossible. They also can pronounce numerous 'axioms' of electoral statistics.

\begin{figure}
\includegraphics[scale=0.5]{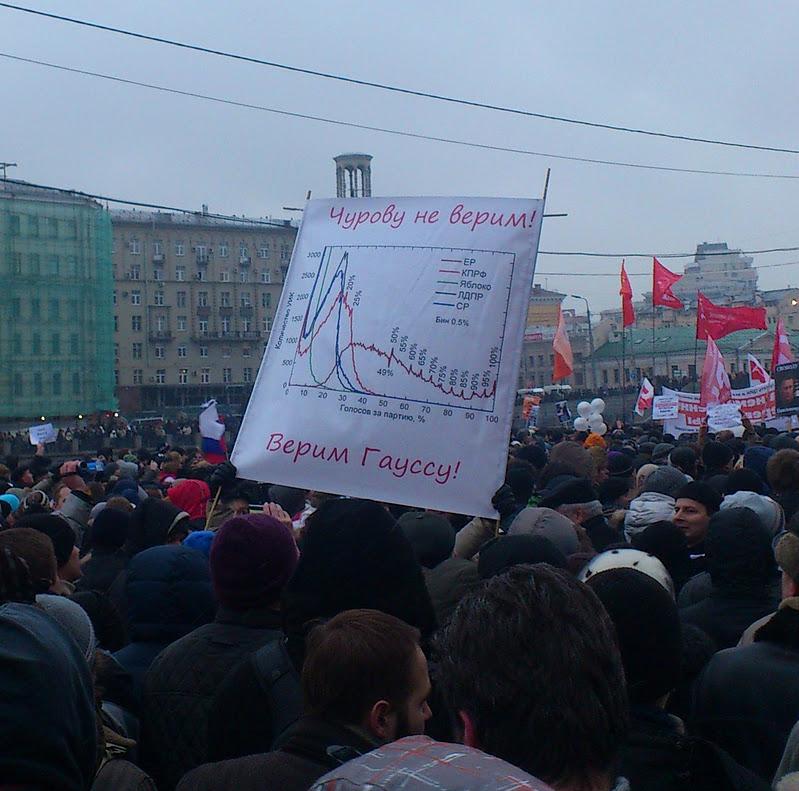}
\caption{Meeting in Moscow 10.12.2011. The graph on the banner is a version of Figure \ref{fig:shp1}.
The text is 'We do not trust Churov, we trust Gauss'. Churov is the chairman
of the Central Electoral Commission.
\label{fig:meeting1}}
\end{figure}

\begin{figure}
\includegraphics[scale=0.5]{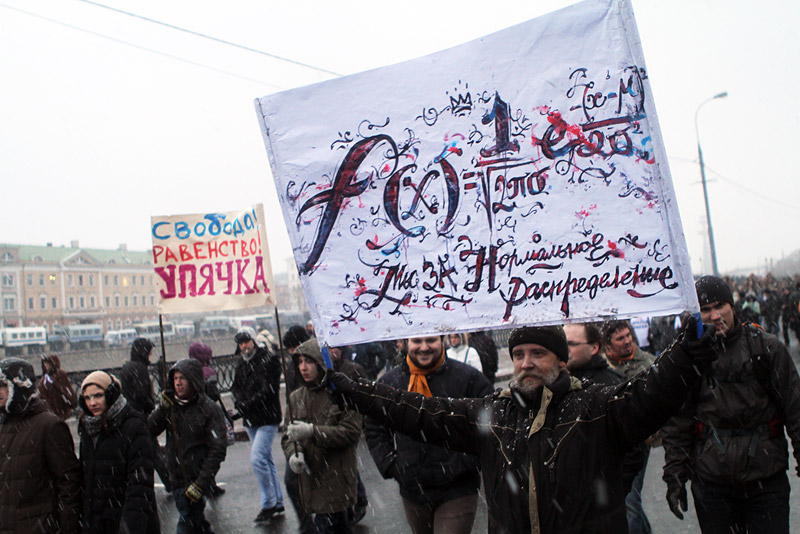}
\caption{Meeting in Sankt-Peterburg, 10.12.2011. The text on the poster is 'For normal distribution!'.
\label{fig:meeting2}}
\end{figure}

\begin{figure}
\includegraphics[scale=0.3]{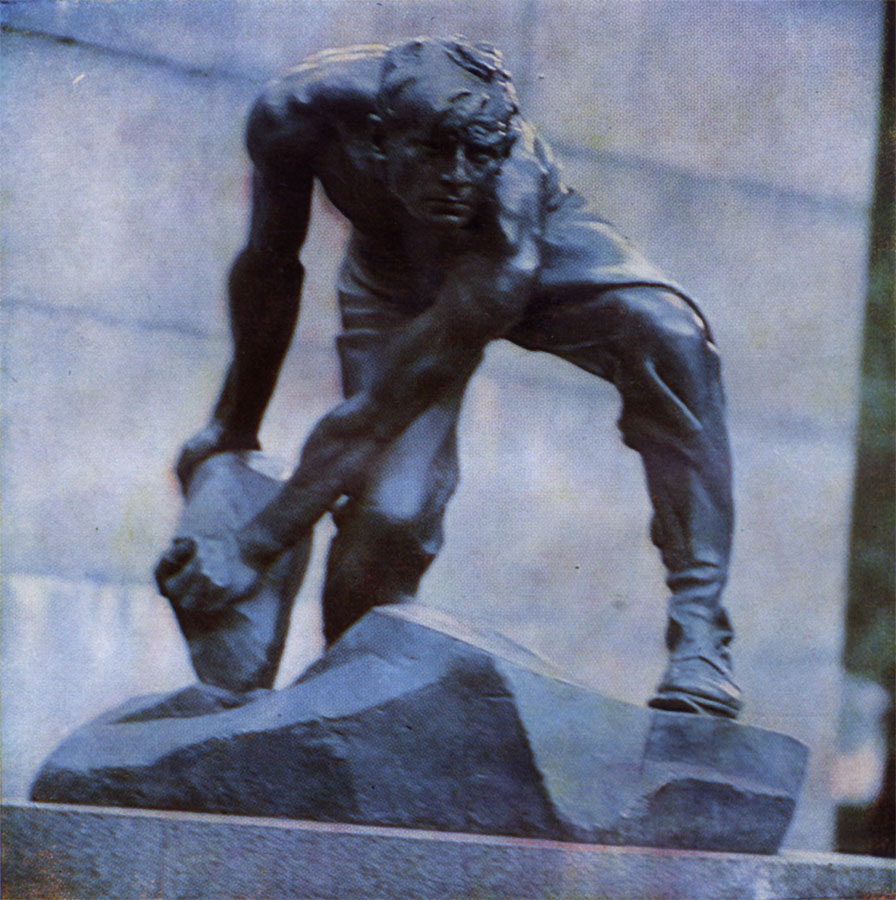}
\caption{Reference to footnote 2 (page 1).
This is the famous sculpture of Ivan Shadr (1927) "Cobblestone is a weapon of proletariat".
\label{fig:shadr}}
\end{figure}

 Certainly this activity will be continued, 
see Figures \ref{fig:meeting1}--\ref{fig:shadr} below after the bibliography.

{\tt Math.Dept., University of Vienna,

 Nordbergstrasse, 15,
Vienna, Austria

\&

Institute for Theoretical and Experimental Physics,

Bolshaya Cheremushkinskaya, 25, Moscow 117259,
Russia

\&

Mech.Math. Dept., Moscow State University,
Vorob'evy Gory, Moscow

e-mail: neretin(at) mccme.ru

URL:www.mat.univie.ac.at/$\sim$neretin

wwwth.itep.ru/$\sim$neretin}

\newpage

{\bf \large Addendum 1. Regional statistics}

\bigskip

Reference to footnote \ref{foot:table}.

Below we present the list of regions of Russian Federation ordered  according percent of voices
for United Russia party (UR).  For each region we present

\sm

--- the total number of electors
(the first column, in brackets, millions);

---  percent of voices for UR (the second column);

---  voting turnout (column 3);

---  the percent of voices for UR
with respect to the total number of electors (column 4);

---  column 5 contains a geographic information (see abbreviations after the table). 

\sm

We distinguish regions according their status in the Constitution.

--- national republics are marked by {\bf bold};

--- ordinary regions (oblasts and krays)
by the usual $\setminus$rm font. 

--- we also mark 4 autonomous okrugs and a unique autonomous oblast
by {\it italic}, apparently these cases must be considered as an exoticism outside
statistics. 

\begin{longtable}{l c c c l l}
1& 2&3&4&5+&5-
\\
\nomer
{\bf Republic of Chechenia} (0.6) & 	99,5 &   99,5 & 	99.0   & \sf NC,I
\\ 
\nomer
{\bf Republic of Mordovia} (0,7) & 	91,6  &  94,2 &   86.3  && \sf For
\\ 
\nomer
{\bf Republic of Daghestan} (1.6) & 	91,4  &   91.1 &  83.3 & \sf NC, I 
\\ 
\nomer
{\bf Republic of Ingushetia} (0.2)& 	91,0 &   86,4  &   78.6 & \sf NC, I
\\ 
\nomer
{\bf 
Karachaevo-Circassian Rep.} (0.3) & 	89,8 &   93,2 &  83.7 & \sf NC, I
\\ 
\nomer
{\bf Republic of Tyva} (0,2) & 	85,3  &  86,1  &  73.4 & &\quad \sf East
\\ 
\nomer
{\bf Rep. of Kabardino-Balkaria} 
 (0.5) &  	81,9 &    98,4 &   80.6 & \sf NC, I
\\ 
\nomer
{\bf Republic of Tatarstan} 
 (2,9) &  	77,8  &   79,5 &  61.9 &  \sf I, \sf Pr
\\ 
\nomer
{\it Yamal-Nenets autonomous okrug}
 (0,4) & 	71,7 &  82,2 &  58.9 & \,\,\sf WS 
\\ 
\nomer
{\bf Republic of Bashkortostan}
 (3,0) & 	70,5 &   79,3 &     55.9 & \sf I & \sf For
\\ 
\nomer
{\it Chukotka autonomous okrug }
 (0,03) & 	70,3  &  79,1  &  55.6 &&\quad \sf East 
\\ 
\nomer
{\bf Republic of North Osetia} 
 (0.5) & 	67,9 &    85,8 &   58.0 & \sf NC
\\

\\\nomer
Tambov oblast
 (0,9) & 	66,7 &    68,3 &  45.6 & \sf Pr
\\ 
\nomer
{\bf Republic of Kalmykia} (0,2) & 	66,1 &  63,2 &  41.8 & \sf Pr 
\\ 
\nomer
Saratov oblast (2,0) & 	64,9 &  67,3  &  43.7 & \sf Pr
\\ 
\nomer
Kemerovo oblast (2,1) & 	64,2 &   69,4 &    44.6 & \,\,\sf WS
\\ 
\nomer
Tyumen oblast
 (1,0) & 	62,2 &   76,2  &    47.4 & \,\,\sf WS
\\ 
\nomer
Tula oblast
 (1,3) & 	61,3 &   72,8  &   44.6 & \sf Pr 
\\ 
\nomer
{\bf Republic of Adygeya}
 (0.3) & 	61,0 &    65,9 &  40.2& \sf NC,I
\\ 
\nomer
Astrakhan oblast
 (0,8) & 	60,2  &  56,0 & 	33.7 & \sf Pr
\\ 
\nomer
{\bf Republic of Komi}
 (0,7) & 	58,8  &  72,6	&   42.7 & & \sf For
\\ 
\nomer
Penza oblast
 (1,1) & 	56,3 &   64,9 &   36.6 & \sf Pr
\\ 
\nomer
Krasnodar kray
 (3.8) & 	56,2 &   72.6 &  40.8 & \sf Pr
\\
\\ 
\nomer

{\bf Republic of Altai}
 (0,2) &  	53,3 &   63,6    &    33.9 & \,\,\sf WS 
\\ 
\nomer
{\bf Republic of Mari}
 (0,5) & 	52,2 &   71,3 &  37.2 & & \sf For
\\ 
\nomer
Belgorod oblast
 (1,2) & 	51,2 &  75,5 &  38.7 & \sf Pr
\\ 
\nomer
Chelyabinsk oblast
 (2,8) & 	50,3 &   59,7  &   30.0  & (\sf Pr)
\\ 
\nomer
Rostov oblast
 (3.3) & 	50,2 &  59.3 &   29.8& \sf Pr 
\\ 
\nomer
Bryansk oblast
 (1,0) &  	50,1 &   59,9 &  30.0 & (\sf Pr) 
\\ 
\nomer
Voronezh oblast
 (1,9) & 	50,0 &   64,3 &  32.2&  \sf Pr
\\ 
\nomer
{\bf Republic of Yakutia}
 (0,6) & 	49,2  &   60,1  &   29.6 &&\quad \sf East 
\\ 
\nomer
Stavropol kray
 (2.0) & 	49,1 &   50.9  &  25.0 &  \sf Pr
\\ 
\nomer
{\bf Republic of Buryatia}
 (0,7) & 	49,0  &  56,9 &   27.9& &\quad \sf East
\\ 
\nomer
{\it Jewish autonomous oblast}
 (0,1) & 	48,1 &  52,1  &   25.0& &\quad \sf East
\\ 
\nomer
MOSCOW-city (7,2) & 	46,6 &   61,7 &   28.9 & &(\sf For)
\\ 
\nomer
Kursk oblast
 (0,9) & 	45,7  &  54,7 &  25.9 & \sf Pr
\\ 
\nomer
Kamchatka kray
 (0,3) & 	45,3  &  53,6   &        24.3 &&\quad \sf East 
\\ 
\nomer
{\bf Republic of Udmurtia}
 (1,2) & 	45,1 &  56,6  & 25.5 & & \sf For
\\ 
\nomer
Nizhny Novgorod oblast
 (2,7) & 	44,6  &  58,9 & 26.3 && \sf For
\\ 
\nomer
Kurgan oblast
 (0,8)&  	44,4 &   56,5&   25.1&  \,\,\sf WS
\\ 
\nomer
Ul'yanovsk oblast
 (1,1) & 	43,6 &  60,4 &   26.3& \sf Pr
\\ 
\nomer
Amur oblast
 (0,7) &  	43,5 &   54,0  &        23.5& &\quad \sf East
\\ 
\nomer
{\bf Republic of Chuvashia}
 (1,0) &  	43,4 &   61,7  &  26.3& & \sf For
\\ 
\nomer
Zabaikal'e kray
 (0,8) & 	43,3 &  53,6 &   23.2 & &\quad \sf East
\\ 
\nomer
Sakhalin oblast
 (0,4) & 	41,9 &  49,1   &     20.6& &\quad \sf East
\\ 
\nomer
Magadan oblast
 (0,1) & 	41,0  &  52,6 &    21.6 & &\quad \sf East
\\ 
\nomer
Kaluga oblast
 (0,8) & 	40,4 &   57,5 &   23.2& & \sf For
\\ 
\nomer
Lipetsk oblast
 (1,0)&  	40,1 &   56,9 &  22.8& \sf Pr
\\ 
\nomer
{\it Khanty-Mansi auton. okrug}
 (1,1) & 	41,0 & 	54,9 & 22.5&  \,\,\sf WS
\\ 
\nomer
{\bf Republic of Khakassia}
 (0,4) & 	40,1  & 56,2& 	22.6& &\quad \sf East
\\ 
\nomer
Ivanovo oblast
 (0,8) & 	40,1  &  53,2 &   21.3& & \sf For
\\ 
\nomer
Ryazan oblast
 (1,0) & 	39,8 &   52,7  &  21.0 & & \sf For
\\ 
\nomer
Omsk oblast
 (1,6) & 	39.6 &  55,7   &      22.1  & \,\,\sf WS 
\\ 
\nomer
Samara oblast
 (2,6) & 	39,4 &  53,0 &   20.9& \sf Pr
\\ 
\nomer
Orel oblast
 (0,7) & 	39,0 &   64,7 &  25.2& \sf Pr
\\ 
\nomer
Tver oblast
 (1,1) & 	38,4 &   53,5 &  20.5& & \sf For
\\ 
\nomer
Vladimir oblast
 (1,3)&  	38,3  &  48,9 & 18.7 & & \sf For
\\ 
\nomer
Khabarovsk kray
 (1,1) &  	38,1 &  53,2 &    20.3 & &\quad \sf East
\\ 
\nomer
Tomsk oblast
 (0,8)&  	37,5 &  50,5 &       18.9&  \,\,\sf WS 
\\ 
\nomer
Altai kray
 (2,0) & 	37,2 &  52,5  &     19.6& \,\,\sf WS
\\ 
\nomer
Kaliningrad oblast
 (0,8) & 	37,1 &   54,6 &  20.3 & & \sf For
\\ 
\nomer
Pskov oblast
 (0,6) & 	36,7 &   52,9 &  19.4 & & \sf For
\\ 
\nomer
Krasnoyarsk kray
 (2,2) & 	36,7 &  49,7  &  18.2& &\quad \sf East
\\ 
\nomer
Perm kray
 (2,1) 	& 36,3&   48,1 &   17.5& & \sf For
\\ 
\nomer
Smolensk oblast
 (0,8) & 	36,2 &   49,6& 	18.0 & & \sf For
\\ 
\nomer
{\it Nenets autonomous okrug}
 (0,0) & 	36,0  &  56,1 &   20.2& & T
\\ 
\nomer
 Volgograd oblast
 (2,0) & 	35,5 &  52,0& 	18.5& \sf Pr
\\ 
\nomer
ST.PETERSBURG (city) (3,6) & 	35,4 &  55,2 &  19.5& &(\sf For)
\\ 
\nomer
Kirov oblast
 (1,1) & 	34,9 &   54,0	&   18.8 & & \sf For
\\ 
\nomer
Orenburg oblast
 (1,6) & 	34,9&   51,2 &  17.7& \sf Pr
\\ 
\nomer
Irkutsk oblast
 (1,9) & 	34,9 &   47,1  &   16.4 & &\quad \sf East
\\ 
\nomer
Novgorod oblast
 (0,5) & 	34,6 &  56,6 &  19.6 & & \sf For
\\ 
\nomer
Novosibirsk oblast
 (2,1)&  	33,8 &  56,8 &  19.2 & \,\,\sf WS
\\ 
\nomer
Vologda oblast
 (1,0) & 	33,4 &   56,3  &  18.8 & & \sf For
\\ 
\nomer
Leningrad oblast
 (1,3) & 	33,0 &   51,5&   17.0 & &\sf For
\\ 
\nomer
Primor'e kray
(1,5)&  	33,0 &  48,7 &    16.1& &\quad\sf \sf East
\\ 
\nomer
Moscow oblast
 (5,6) & 	32,8 &  51,0 &  16.7 & &\sf For
\\ 
\nomer
Sverdlovsk oblast
 (3,5)&  	32,7 &  51,2 &   16.7 &&\sf For 
\\ 
\nomer
{\bf Republic of Karelia} (0,6) & 	32,3 &   50,3 &   16.2 &&\sf For 
\\ 
\nomer
Murmansk oblast
 (0,7) & 	32,0 &   51,8 &   16.6 & & \sf For, T
\\ 
\nomer
Archangelsk oblast
 (1,0) & 	31,9 &   50,0 &  16.0& & \sf For
\\ 
\nomer
Kostroma oblast
 (0,6) & 	30,7 &  57,3 &  17.6& &\sf For
\\ 
\nomer
Yaroslavl oblast
 (1,1) & 	29,0 &   55,9 &  16.2& &\sf For
\end{longtable}

We see that voting in 'republics' and ordinary regions was strongly different,
but 'republics' also are different. We  observe impressive 12 top-lines of the table
(recall that autonomous {\it okrugs} were omitted from the considerations).

We split the table on 3 parts, lines 1--12, 13--23, 24--83 respectively.
 The first and the second groups 
are distinguished due a jump in the 4th column of the table  
61.9\%--(58.9\%)--58.0\%--(55.9\%)--55.6\% in the leading group
and 45.6\% (Tambov), 47.6\% (Tyumen).

The boundary between the second and the third group is not precise. Here there is a jump
56.2\%--53.3\% in the second column (but attributions of Astrakhan, Mari,
 and Belgorod can be regarded
as problematic). 

The last group 24-83 is a dense. 

\sm

Abbreviations in Column 5.

{\sf NC} --- North Caucasus;

{\sf I} --- a republic with Islamic title nation;

{\sf Pr} --- a region in European prairies or forest-steppe (the south of European part
of Russia); 

{\sf For} --- a region in European forest zone (the north of European part of Russia); 

{\sf T} --- a region in European tundra (Arctic prairie);

{\sf WS} ---  West Siberia;

{\sf East} --- East Siberia and Far East (Pacific).

\newpage

{\bf \large Addendum 2. Voting stations with given results of United Russia.}

\begin{figure}
\includegraphics[scale=0.4]{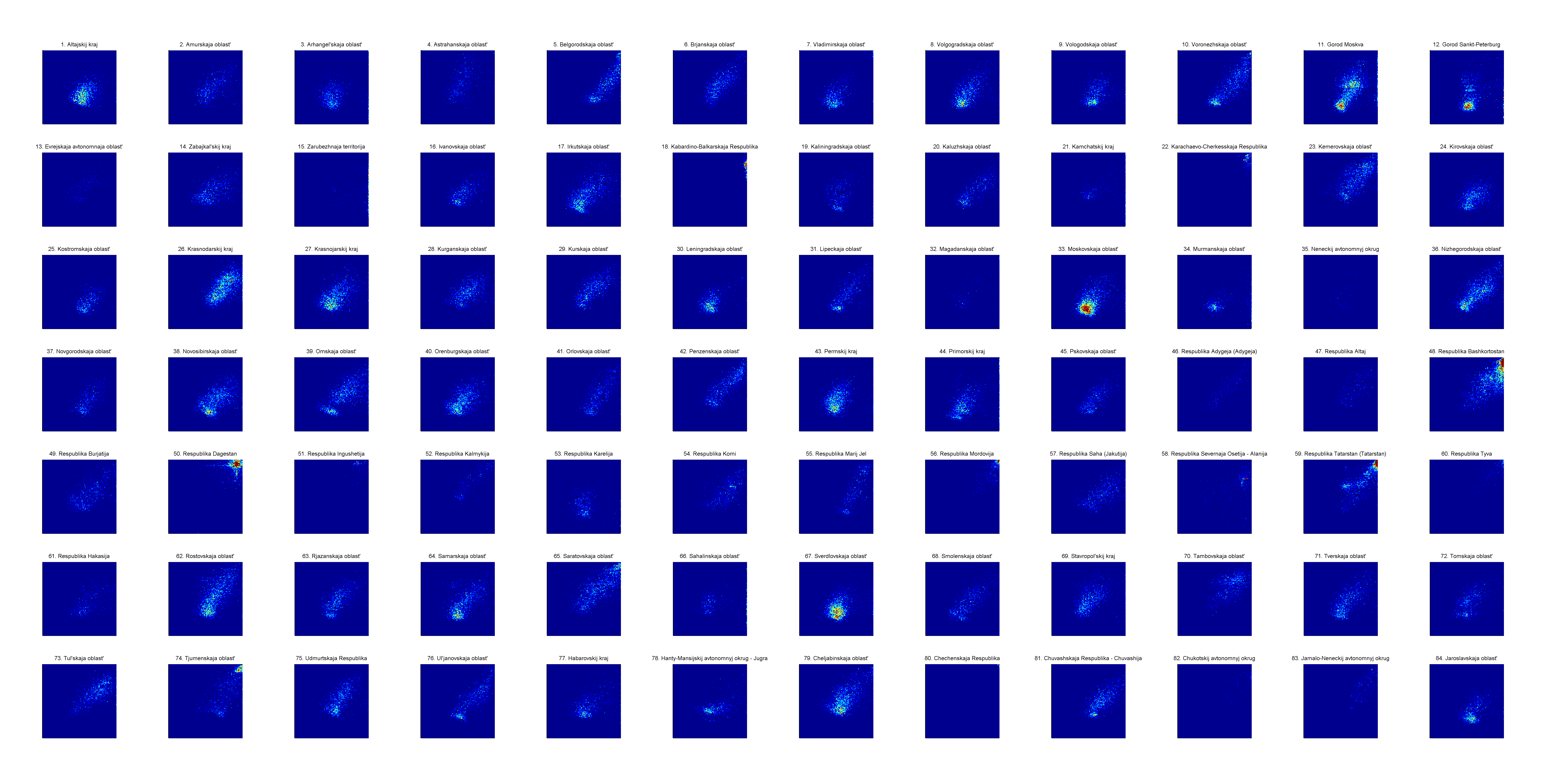}
\caption{(Kobak, http://kobak.610.ru/lj/elections10big.png) 
ER-clouds for all regions of Russian Federation.
Figures are observable after zoom.
\label{fig:vsevse}}
\end{figure}

Reference to footnote \ref{foot2}.

Figure \ref{fig:vsevse} presents UR-cloud diagrams for all regions of Russia.
For ordinary regions pictures looks as comets, sometimes ellipses or circles.
 Moscow-city discussed above is slightly  unusual
(oval with waist). Separately, we present  cloud diagrams for the exceptional regions.
It seems that they are very individual and very strange. It seems that elements of real elections
in the Republics of Bashkortostan, Tatarstan, and North Osetia are visible on these Figures.

\newpage

\end{document}